\documentclass[10pt,journal,twocolumn]{IEEEtran}
\usepackage[utf8]{inputenc}
\usepackage{amsthm}
\usepackage{amsfonts}
\usepackage{amssymb}
\usepackage{graphicx}
\usepackage{epstopdf}
\renewcommand{\vec}[1]{\mathbf{#1}}
\usepackage[cmex10]{amsmath}
\usepackage[justification=centering]{caption}
\usepackage{enumerate} 
\usepackage{cite}
\usepackage{color}
\usepackage[dvipsnames]{xcolor}
\newcommand{\vast}{\bBigg@{4}}
\newcommand{\Vast}{\bBigg@{5}}

\usepackage{suffix}
\usepackage{mathtools}
\usepackage[list=true]{subcaption}

\DeclarePairedDelimiterX\MeijerM[3]{\lparen}{\rparen}%
{\begin{smallmatrix}#1 \\ #2\end{smallmatrix}\delimsize\vert\,#3}

\newcommand\MeijerG[8][]{%
  G^{\,#2,#3}_{#4,#5}\MeijerM[#1]{#6}{#7}{#8}}

\WithSuffix\newcommand\MeijerG*[7]{%
  G^{\,#1,#2}_{#3,#4}\MeijerM*{#5}{#6}{#7}}
  
  \DeclarePairedDelimiterX\FM[7]{\lparen}{\rparen}{\begin{smallmatrix}#1  \end{smallmatrix}\delimsize\vert\,\begin{smallmatrix}#2 \\ #3 \\ #4 \\ #5 \end{smallmatrix} \delimsize\vert\,\begin{smallmatrix}#6 \\ #7  \end{smallmatrix}}

\begin{document}

\title{Analysis of Outage Probability of MRC with $\eta-\mu$ co-channel interference}

\ifCLASSOPTIONtwocolumn
\author{Muralikrishnan Srinivasan, Sheetal Kalyani\\
\hspace{-0.5 cm}Department of Electrical Engineering,\\
  \hspace{-1cm} Indian Institute of Technology, Madras, \\
 \hspace{-1cm} Chennai, India 600036.\\
 \hspace{-1.5cm} \{ee14d206,skalyani\}@ee.iitm.ac.in}   
\else
\author{Muralikrishnan Srinivasan  \hspace{1cm} Sheetal Kalyani   \\
 \hspace{-0.5 cm}Department of Electrical Engineering,\\
  \hspace{-1cm} Indian Institute of Technology, Madras, \\
 \hspace{-1cm} Chennai, India 600036.\\
 \hspace{-1.5cm} \{ee14d206,skalyani\}@ee.iitm.ac.in\\
 }
 \fi
 \maketitle
 
 \begin{abstract}
Approximate outage probability expressions are derived for systems employing maximum ratio combining, when both the desired signal and the interfering signals are subjected to $\eta-\mu$ fading, with the interferers having unequal power. The approximations are in terms of the Appell Function and Gauss hypergeometric function. A close match is observed between the outage probability result obtained through the derived analytical expression and the one obtained through Monte-Carlo simulations.
\end{abstract}

\section{Introduction}
Various diversity combining schemes have been proposed to exploit space diversity offered by adaptive antenna arrays. Maximum ratio combining (MRC) is one such scheme that maximizes the signal to noise Ratio (SNR) at the output of the receiver. Though, in the presence of co-channel interference (CCI), the performance of MRC is sub-optimal, MRC is still widely used as it has lower receiver complexity when compared with optimum combining (OC).
\par Outage probability ($P_{out}$) is an important measure of wireless system performance. Outage probability of MRC has been studied in the presence of CCI for Rayleigh channels in \cite{shah_ray_mrc, zhang_corr_mrc} and Rician channels in \cite{chayawan}. Outage probability analysis of MRC has been studied for Nakagami faded channels in \cite{jerez_nakagami_mrc} with the assumption that the interferer fading is the same in all diversity branches and in \cite{wang_nakagami_mrc} with no such assumption.
\par Recently, there has been significant focus on generalized fading models namely $\kappa-\mu$ and $\eta-\mu$ models introduced in \cite{yacoub_k_mu}. These distributions model the small-scale variations in the fading channel in the line of sight and non-line of sight conditions respectively. Further, these generalized fading distributions include Rayleigh, Rician, Nakagami, One-sided Gaussian distributions as special cases. 
\par There has been some work in literature, for example \cite{ansari_mrc, morales_nmu_mrc, paris_nmu_mrc,chen_outage}, which analyze the performance of MRC, in the presence of these generalized faded signals. In \cite{morales_nmu_mrc, chen_outage}, generalized fading channels are assumed only for the signal of interest (SOI), whereas interferers are assumed to be Rayleigh faded. In \cite{ansari_mrc}, the SOI is assumed to be $\eta-\mu$ distributed, but CCI is not considered. Extending the analysis in \cite{morales_nmu_mrc, chen_outage} for the case of general $\eta-\mu$ interferers seems mathematically intractable, if not impossible. Outage probability expressions, when interferers are assumed to be $\eta-\mu$ faded, were first derived in \cite{paris_nmu_mrc}. However, in \cite{paris_nmu_mrc}, the interference is assumed to have the same realization across the diversity branches, while the user signal has different realization across the diversity branches, which limits its practical utility. Hence, in this work, we determine an approximate expression for outage probability when both SOI and the CCI are $\eta-\mu$ distributed, with the interference fading being different across diversity branches. We assume that the noise power is negligible when compared to the interference power, as is the case in most practical wireless systems. The interferers are assumed to be uncorrelated and have unequal powers. To the best of our knowledge, ours is the first work in open literature, to give at least an approximate expression for the outage probability of MRC when both the user and the unequal power interferers are assumed to be $\eta-\mu$ distributed.
\section{System model}
Let the number of receive antennas be $N_R$ and let the number of interferers be $N_I$. The $N_R \times 1$ user signal $\vec c = [ c[1],...,c[N_R]]^T$ are assumed to be $\eta-\mu$ random variables \cite{eta_mu_phase} and the $N_I$ interferers denoted $\vec c_i$ are also independent $\eta -\mu$  random variables with complex probability density function (pdf) given by \cite{eta_mu_phase}
\begin{align}\label{nmu}
f(x, y) &= \frac{\mu ^{2 \mu}|xy|^{2\mu-1}}{\Omega_X^{\mu} \Omega_Y^{\mu} \Gamma^2(\mu)}exp[-\mu\Big(\frac{x^2}{\Omega_X}+\frac{y^2}{\Omega_Y}\Big)],
\end{align}
where $\Omega$ is the power parameter given by $\Omega=2 \sigma^2 \mu$. Here, $\sigma^2$ is the power of the Gaussian variable in each cluster and $2 \mu$ is the number of clusters. Note, $\Omega_X=(1-\eta)\Omega/2$, $\Omega_Y=(1+\eta)\Omega/2$ and $-1 \leq \eta \leq 1$ (Format 2). Henceforth, the subscript $int$ and $user$ will be used to differentiate interferer $\eta-\mu$ parameters from user $\eta-\mu$ parameters.
$u$ denotes the desired user symbol and $u_i$  the $i^{th}$ interferer symbol, with both belonging to unit energy QAM constellation. The $N_R \times 1$ received vector in a receive diversity system is given by
\begin{equation}\label{R}
\vec r =\vec c u + \sum_{i=1}^{N_I} \sqrt{E_i}\vec c_i u_i ,
\end{equation} 
where $E_i$ is the mean energy of each of the interfering signals.
In MRC, the received vector $\vec r$ is weighted by $\vec c^H$ to produce the following output at the combiner:
\begin{equation}
z =\vec c^H \vec r=\vec c^H \vec c u +\sum_{i=1}^{N_I} \sqrt{E_i} \vec c^H \vec c_i u_i.
\end{equation}
Hence the SIR at the output of the combiner is given by \cite{shah_ray_mrc}
\begin{equation}\label{mrcsir1}
\gamma= \frac{ |\vec c^H \vec c|^2 }{\sum_{i=1}^{N_I} E_i | \vec c^H \vec c_i|^2} = \frac{ |\vec c^H \vec c|^2 }{\sum_{i=1}^{N_I} E_i  \vec c^H \vec c_i \vec c_i^H \vec c}.
\end{equation}
\section{Preliminaries}
Obtaining the exact pdf of $\gamma$ seems mathematically intractable. Hence, we go in for a simple approximation to obtain the pdf. Consider the expression for SIR given by (\ref{mrcsir1}). We propose to approximate each of the term in the denominator $E_i | \vec c^H \vec c_i|^2$ by $x_i$, which is defined by the product of $|\vec c^H \vec c|$ and a gamma random variable $z_i$ ( independent of $|\vec c^H \vec c|$), for $ i=1,...,N_I$, by matching the first and the second moment. In other words, we match the first and the second moment of $E_i | \vec c^H \vec c_i|^2$ and $x_i= |\vec c^H \vec c| z_i$.\footnote{This method is inspired from the case of Rayleigh faded inteferers \cite{shah_ray_mrc}, where each of the term in the denominator is exactly a product of $ |\vec c^H \vec c|$ and a gamma random variable.}
\par The gamma random variable $z_i \sim \mathcal{G}(a_i, b_i)$ is given by its pdf  
$f_{z_i}(z_i)= \frac{1}{\Gamma(a_i) b_i^{a_i} }{z_i}^{a_i-1} e^{-\frac{z_i}{b_i}}$,  
with $z_i \in (0, \infty)$, $a_i >0 $ being the shape parameter and $b_i >0$ being the scale parameter. 
The first moment of the term $ | \vec c^H \vec c_i|^2$ is given by \cite{yacoub_k_mu},
\begin{equation}\label{term1moment}
\mathbb E[E_i |\vec c^H \vec c_i|^2] = E_i N_R ( \Omega_{Xuser} + \Omega_{Yuser})(\Omega_{Xint} + \Omega_{Yint})
\end{equation} 
and the second moment can be derived by expanding $ |\vec c^H \vec c_i|^2$, taking its square and taking expectation of each term explicitly, to obtain,
\ifCLASSOPTIONtwocolumn
\begin{multline}\label{term2moment}
\mathbb E[(E_i |\vec c^H \vec c_i|^2)^2] = E_i^2 \Bigg[ \Big[( \Omega_{Xuser}^2 + \Omega_{Yuser}^2)(\Omega_{Xint}^2 + \Omega_{Yint}^2)\\
\times \Big(3 N_R (N_R-1)+ N_R \frac{\mu_{user}+1}{\mu_{user}}\frac{\mu_{int}+1}{\mu_{int}} \Big)\Big]\\
 +\Big[ ( \Omega_{Xuser}^2 + \Omega_{Yuser}^2)(\Omega_{Xint}  \Omega_{Yint}) \Big( 2 N_R (N_R-1)\\
 +2 N_R \frac{\mu_{user}+1}{\mu_{user}} \Big)\Big]
 + \Big[( \Omega_{Xint}^2 + \Omega_{Yint}^2)(\Omega_{Xuser}  \Omega_{Yuser})\\
 \times \Big( 2 N_R (N_R-1)+2 N_R \frac{\mu_{int}+1}{\mu_{int}} \Big)\Big]
 + \Big[(\Omega_{Xint}  \Omega_{Yint}\\ \times \Omega_{Xuser}  \Omega_{Yuser})
\Big( 4 N_R^2 + 8 N_R (N_R-1) \Big)\Big]\Bigg],
\end{multline}
\else
\begin{align}\label{term2moment}
\nonumber
\mathbb E[(E_i |\vec c^H \vec c_i|^2)^2] &= E_i^2 \Bigg[( \Omega_{Xuser}^2 + \Omega_{Yuser}^2)(\Omega_{Xint}^2 + \Omega_{Yint}^2) \Big(3 N_R (N_R-1)+ N_R \frac{\mu_{user}+1}{\mu_{user}}\frac{\mu_{int}+1}{\mu_{int}} \Big)\\
\nonumber
& \quad + ( \Omega_{Xuser}^2 + \Omega_{Yuser}^2)(\Omega_{Xint}  \Omega_{Yint}) \Big( 2 N_R (N_R-1)+2 N_R \frac{\mu_{user}+1}{\mu_{user}} \Big)\\
\nonumber
& \qquad + ( \Omega_{Xint}^2 + \Omega_{Yint}^2)(\Omega_{Xuser}  \Omega_{Yuser}) \Big( 2 N_R (N_R-1)+2 N_R \frac{\mu_{int}+1}{\mu_{int}} \Big)\\
& \quad \qquad + (\Omega_{Xint}  \Omega_{Yint} \Omega_{Xuser}  \Omega_{Yuser}) \Big( 4 N_R^2 + 8 N_R (N_R-1) \Big)\Bigg],
\end{align}
\fi
where $\mathbb E[.] $ denotes expectation \footnote{A similar expression can be found for i.n.i.d. $\eta-\mu$ interferers, but not included here.}.
The first moment of gamma random variable is given by $\mathbb E[z_i]= a_i b_i$. Hence, matching the first moment is equivalent to
\begin{equation}\label{matchone}
\mathbb E[x_i]= E[|\vec c^H \vec c|] a_i b_i,
\end{equation}
where $E[|\vec c^H \vec c|]=N_R ( \Omega_{Xuser}+ \Omega_{Yuser})$.
The second moment is given by $\mathbb E[z_i^2]= a_i b_i^2 + a_i^2 b_i^2$. Hence, matching the second moment is equivalent to,
\ifCLASSOPTIONtwocolumn
\begin{align}\label{matchtwo}
\mathbb E[x_i^2]&=E[|\vec c^H \vec c|^2] (a_i b_i^2 + a_i^2 b_i^2),
\end{align}
\else
\begin{equation}\label{matchtwo}
\mathbb E[x_i^2]= E[|\vec c^H \vec c|^2] (a_i b_i^2 + a_i^2 b_i^2),
\end{equation}
\fi
where $E[|\vec c^H \vec c|^2]=N_R \Big[ (\frac{\mu_{user}+1}{\mu_{user}}+ N_R -1)(\Omega_{Xuser}^2+ \Omega_{Yuser}^2)+ 2 N_R \Omega_{Xuser} \Omega_{Yuser} \Big]$. Note that $E[|\vec c^H \vec c|^2]$ can be obtained by expanding the square and explicitly calculating the expectation of each term.  
From (\ref{matchone}) and (\ref{matchtwo}), we obtain,\\
$b_i= \frac{ \frac{\mathbb E[(E_i |\vec c^H \vec c_i|^2)^2]}{\mathbb E[|\vec c^H \vec c|^2]}  - \Bigg(\frac{\mathbb E[E_i |\vec c^H \vec c_i|^2]}{\mathbb E[|\vec c^H \vec c|]} \Bigg)^2 }{ \frac{\mathbb E[E_i |\vec c^H \vec c_i|^2] }{\mathbb E[|\vec c^H \vec c|]}}$ and 
$a_i=\frac{\frac{\mathbb E[E_i |\vec c^H \vec c_i|^2] }{\mathbb E[|\vec c^H \vec c|]}}{b_i}$.
Now, the SIR at the output of the combiner is given by,
\begin{equation}\label{mrcsir3}
\gamma \approx \frac{ |\vec c^H \vec c|^2 }{ \sum_{i=1}^{N_I} |\vec c^H \vec c| z_i} = \frac{ |\vec c^H \vec c|^2 }{|\vec c^H \vec c| \sum_{i=1}^{N_I}  z_i} =\frac{ |\vec c^H \vec c| }{ \sum_{i=1}^{N_I}  z_i}.
\end{equation}
The denominator is now a sum of $N_I$ gamma random variables $z_i$. Also, from \cite{heath_poisson},  it is again known that a sum of gamma variables can be well approximated by a single gamma random variable by matching the first two moments Hence approximately, $\sum_{i=1}^{N_I}z_i \sim \mathcal{G}(a,b)$,
where 
\begin{equation}\label{a}
 a= \frac{\big(\sum_{i=1}^{N_I}a_i b_i \big)^2}{\sum_{i=1}^{N_I}a_i b_i^2}
\end{equation} and 
\begin{equation}\label{b}
b= \frac{\sum_{i=1}^{N_I}a_i b_i^2}{ \sum_{i=1}^{N_I}a_i b_i}.
\end{equation}
Note that for $E_1= E_2=...= E_{N_I}$, the above approximation of sum of gamma random variables by another gamma random variable becomes exact.
\par  The SIR at the output of the combiner is given by 
$\gamma \approx \frac{y}{x}$,
where $y$ is the sum of $N_R$ $\eta-\mu$ power variables denoted by $y_i$ $i=1,..,N_R$ and $x \sim \mathcal{G}(a,b)$. 
Two different approximations are possible depending on whether the numerator is also approximated by a gamma random variate. 
\section{Approximations}
\subsection*{Approximation 1}
The pdf of sum of $N_R$ i.i.d. $\eta-\mu$ random variables is given by \cite{yacoub_k_mu},
\ifCLASSOPTIONtwocolumn
\begin{align}\label{norm_nmu}
\nonumber
f_{Y}(y)
&=\frac{2 \sqrt{\pi}   (N_R \mu)^{2 N_R \mu} h^{N_R \mu}}{ \Gamma(N_R \mu) \Gamma(N_R \mu+1/2) \bar y} (y/ \bar y)^{2 N_R \mu-1} e^{-2 N_R \mu h y/ \bar y}\\ &\quad  \,_0F_1(N_R  \mu+1/2, N_R^2\mu^2 H^2 y^2/ \bar y^2),
\end{align} 
\else
\begin{align}\label{norm_nmu}
f_{Y}(y)
&=\frac{2 \sqrt{\pi}   (N_R \mu)^{2 N_R \mu} h^{N_R \mu}}{ \Gamma(N_R \mu) \Gamma(N_R \mu+1/2) \bar y} (y/ \bar y)^{2 N_R \mu-1} e^{-2 N_R \mu h y/ \bar y}  \,_0F_1(N_R  \mu+1/2, N_R^2\mu^2 H^2 y^2/ \bar y^2),
\end{align} 
\fi
where $\bar y =N_R E[ y_i]=N_R ( \Omega_{Xuser}+ \Omega_{Yuser})$.
The pdf of SIR $\gamma$ which is the ratio of two positive random variables is given by,
\begin{equation}
f_{\gamma}(\gamma)= \int_{0}^{\infty} x f_Y(\gamma x) f_X(x) dx.
\end{equation}
Substituting the respective pdfs in the above expression and solving the above integration using \cite[Eq 7.525]{int}, we obtain,
\ifCLASSOPTIONtwocolumn
\begin{align}
\nonumber
f_{\gamma}(\gamma) &\approx \frac{2 \sqrt{\pi}  (N_R \mu_{user})^{2 N_R \mu_{user}} h^{N_R \mu_{user}}  \Gamma(2 N_R \mu_{user} +a)}{ \Gamma(N_R \mu_{user}) \Gamma(N_R \mu_{user}+1/2) \Gamma(a) b^a \gamma}   \\
\nonumber
 &\times \big(\frac{\gamma}{\bar y}\big)^{2 N_R \mu_{user}} \big(2 N_R \mu_{user} h \frac{\gamma}{\bar y} +\frac{1}{b} \big)^{-2 N_R \mu_{user} -a}\\
 \nonumber
 & \times \,_2F_1\big( \frac{2 N_R \mu_{user}+a}{2}, \frac{2 N_R \mu_{user}+a+1}{2}, \\
 & \quad N_R  \mu_{user}+\frac{1}{2},  \frac{4 N_R^2 \mu_{user}^2 H^2  \frac{\gamma^2}{\bar y^2}}{(2 N_R \mu_{user} h \frac{\gamma}{\bar y} +\frac{1}{b})^2} \big).
\end{align}
\else
\begin{align}
\nonumber
f_{\gamma}(\gamma) &\approx \frac{2 \sqrt{\pi}  (N_R \mu_{user})^{2 N_R \mu_{user}} h^{N_R \mu_{user}}  \Gamma(2 N_R \mu_{user} +a)}{ \Gamma(N_R \mu_{user}) \Gamma(N_R \mu_{user}+1/2) \Gamma(a) b^a \gamma} \big(\frac{\gamma}{\bar y}\big)^{2 N_R \mu_{user}}  \big(2 N_R \mu_{user} h \frac{\gamma}{\bar y} +\frac{1}{b} \big)^{-2 N_R \mu_{user} -a}\\
 & \quad \times \,_2F_1\big( \frac{2 N_R \mu_{user}+a}{2}, 
  \frac{2 N_R \mu_{user}+a+1}{2}, N_R  \mu_{user}+\frac{1}{2}, 4 \frac{N_R^2 \mu_{user}^2 H^2  \frac{\gamma^2}{\bar y^2}}{(2 N_R \mu_{user} h \frac{\gamma}{\bar y} +\frac{1}{b})^2} \big).
\end{align}
\fi
Determining outage probability by integrating the above expression is mathematically intractable. Hence, we propose an alternate method. We know that the outage probability is given by $ P_{out} = P[ \gamma < \gamma_0]  \approx P[\frac{y}{ x}< \gamma_0]= 1-P[x <  \frac{y }{\gamma_0}]$,
 where $\gamma_0$ is the target SIR.
 The Cumulative Distribution Function (CDF) of Gamma random variable is given by \cite{suman_outage}, $F_X(x)= \frac{(x/b)^a}{\Gamma(a+1)}\,_1F_1(a,a+1,-x/b)$. Hence the outage probability is given by
 \begin{align*}
 P_{out} &\approx 1-\mathbb{E}_y\big(\frac{(\frac{y}{b \gamma_0})^a}{\Gamma(a+1)}\,_1F_1(a,a+1,\frac{-y}{ b\gamma_0})\big), 
 \end{align*}
where $\mathbb{E}_y$ denotes expectation with respect to the $\eta-\mu$ random variable $y$. 
Hence, $P_{out}$ is given by 
\begin{align*}
 P_{out} &\approx 1- \int_{0}^{\infty}\frac{(\frac{y}{b \gamma_0})^a (\frac{y}{ \bar y})^{2 N_R \mu-1}}{\Gamma(a+1)}\,_1F_1(a,a+1,\frac{-y}{b \gamma_0}) f(y)dy,
\end{align*}
where $f(y)$ is given by (\ref{norm_nmu}). The $\,_0F_1$ Hypergeometric function in (\ref{norm_nmu}) can be first replaced by $\,_1F_1$ Hypergeometric function using the identity $\,_0F_1(b,z)=e^{-2 \sqrt{z}} \,_1F_1(b-\frac{1}{2}, 2b-1, 4\sqrt{z})$ from \cite{abr}.
We then do a change of variable $t= 2 N_R \mu (h+H)y/ \bar y$ and then simplify using the integral identity $F_2(a,b,b';c,c';w,z)=\frac{1}{\Gamma(a)}\int_{0}^{\infty}x^{a-1}e^{-x}\,_1F_1(b;c;wx) \,_1F_1(b';c';zx) dx$ from \cite[Eq. 26]{appell} to obtain the outage probability as
\ifCLASSOPTIONtwocolumn
\begin{align*}
 P_{out} &\approx 1- A \frac{\Gamma(2 N_R \mu+a)}{(2 N_R \mu (h+H))^{2 N_R \mu +a}}\,F_2 \big(2 N_R \mu +a, a, N_R \mu; \\
 &\quad a+1, 2 N_R  \mu; \frac{ -\bar y}{2 N_R \mu (h+H) b \gamma_0} , \frac{2  H}{h+H} \big),
\end{align*}
\else
\begin{align*}
 P_{out} &\approx1- A \frac{\Gamma(2 N_R \mu+a)}{(2 N_R \mu (h+H))^{2 N_R \mu +a}} \\
 &\times   \,F_2 \big(2 N_R \mu +a, a, N_R \mu; a+1, 2 N_R  \mu; \frac{ -\bar y}{2 N_R \mu (h+H) b \gamma_0} , \frac{2  H}{h+H} \big),
\end{align*}
\fi
where $A= \frac{(\frac{ \bar y}{b\gamma_0})^a}{\Gamma(a+1)} \frac{2 \sqrt{\pi}   (N_R \mu)^{2 N_R \mu} h^{N_R \mu}}{ \Gamma(N_R \mu) \Gamma(N_R \mu+1/2)}$ and $F_2(.)$ is the Appell function \cite{appell}  .
The series expansion for  $F_2(a,b,b';c,c';w,z)$ converges only if $|w|+|z| <1$. Hence, using the identity  $F_2(a,b,b';c,c';w,z)= (1-w)^{-a} F_2(a, c-b, b'; c, c', \frac{w}{w-1}, \frac{z}{1-w})$, outage probability can be simplified as
\ifCLASSOPTIONtwocolumn
\begin{align}\label{Pout}
\nonumber
 P_{out} &\approx 1- A \frac{\Gamma(2 N_R \mu+a)}{(2 N_R \mu (h+H) + (\frac{ \bar y}{b \gamma_0}))^{2 N_R \mu +a}} \\
 \nonumber
 &\times   \,F_2 \big(2 N_R \mu +a, 1, N_R \mu; a+1, 2 N_R  \mu; \\
 & \: \frac{(\frac{ \bar y}{b \gamma_0})}{2 N_R \mu (h+H)+ (\frac{ \bar y}{ b \gamma_0})} , \frac{4 N_R \mu  H}{ 2 N_R \mu (h+H)+ (\frac{ \bar y}{b \gamma_0})} \big).
\end{align}
\else
\begin{align}\label{Pout}
\nonumber
 P_{out} &\approx 1- A \frac{\Gamma(2 N_R \mu+a)}{(2 N_R \mu (h+H) + (\frac{ \bar y}{b \gamma_0}))^{2 N_R \mu +a}} \\
 &\times   \,F_2 \big(2 N_R \mu +a, 1, N_R \mu; a+1, 2 N_R  \mu; \frac{(\frac{ \bar y}{b \gamma_0})}{2 N_R \mu (h+H)+ (\frac{ \bar y}{ b \gamma_0})} , \frac{4 N_R \mu  H}{ 2 N_R \mu (h+H)+ (\frac{ \bar y}{b \gamma_0})} \big).
\end{align}
\fi
The outage probability can now be evaluated using either the series expansion \cite[eq. (82)]{appell} or an integral expression for $F_2(.)$ \cite[eq. (20)]{appell}. Also, the approximation is exact when the interferers are subject to Rayleigh fading which is a special case of $\eta-\mu$ fading that is obtained by substituting $\eta=0$ and $2 \mu=1$.
\subsection*{Approximation 2}
Alternatively, the $\eta-\mu$ random variable in the numerator can also be approximated by a Gamma random variable $y' \sim  \mathcal{G}(p, q)= q \mathcal{G}(p, 1) $ using moment matching. Equating with the moments of $y'$, we obtain $p=\frac{2 \mu_{user} N_R}{1+\eta_{user}^2}$ and $q= \frac{N_R(\Omega_{X_{user}}+\Omega_{Y_{user}})}{p}$. Hence 
$\gamma \approx \frac{y'}{x}= \frac{q}{b} \frac{ \mathcal{G}(p, 1)}{ \mathcal{G}(a, 1)}$. The ratio of two gamma distribution $z= \frac{ \mathcal{G}(p, 1)}{ \mathcal{G}(a, 1)}$ is called the beta-prime distribution or beta distribution of the second kind with pdf given by $\frac{\Gamma(a+p)}{\Gamma(a)\Gamma(p)}z^{p-1}(1+z)^{-a-p}$ for $z >0$ \cite{dubey}.
Hence the pdf of SIR can also be given as . 
\begin{align*}
f_{\gamma}(\gamma) \approx \frac{\Gamma(a+p)}{\Gamma(a)\Gamma(p)}(\frac{b}{q})^p\gamma^{p-1}(1+\frac{b}{q}\gamma)^{-a-p} \quad \gamma >0
\end{align*}
 The CDF of a beta-prime distributed random variable, say $Z$, with parameters $m$ and $n$, is given in \cite[Eq 2]{cordeiro}, as $P(Z <z) = \frac{(n)_{m} z^{m}  \,_2F_1(m+n, m, m+1, -z)}{  \Gamma(m+1)}$ for $z \geq 0$.
Therefore, outage is given by,
 \begin{align}\label{approx2}
P_{out}  \approx  \frac{(a)_{p} (\frac{b}{q}\gamma_0)^{p} \,_2F_1(p+a, p, p+1, \frac{-b}{q}\gamma_0)}{  \Gamma(p+1)},
\end{align}
 This approximation involves only a simple Gauss hypergeometric function. We can make the following inferences from (\ref{approx2}):
\begin{itemize}
\item[I1)] Since the approximate outage probability given by (\ref{approx2}) is a CDF evaluated at $\frac{b}{q}\gamma_0$, $P_0(\frac{b}{q}\gamma_0+\delta) > P_0(\frac{b}{q}\gamma_0)$ for $\delta >0$. In other words, decrease in user power $\Omega_{X_{user}}+\Omega_{Y_{user}}$, which corresponds to a decrease in $q$ or increase in target SIR $\gamma_0$ increases outage probability. 
\item[I2)] From \cite[Theorem 1.A.12]{shaked} and the pdf of beta-prime distributed random variables given in \cite{dubey}, for two different sets of parameters namely $(p,a)$ and $(p, a+\delta)$, one can prove that 
\begin{align*}
&\mathcal{S}^-\Big(\frac{\Gamma(a+p)}{\Gamma(a)\Gamma(p)}z^{p-1}(1+z)^{-a-p}\\
&-\frac{\Gamma(a+p+\delta)}{\Gamma(p)\Gamma(a+\delta)}z^{p-1}(1+z)^{-a-p-\delta}\Big)=1,
\end{align*}
where $\mathcal{S}^-$ denotes the number of sign changes and $\delta >0$. The sign change from $-$ to $+$ occurs at $z= \Big(\frac{(a+\delta)_{p}}{(a)_{p}}\Big)^{\frac{1}{\delta}}-1$. This implies that, using the definition of stochastic ordering from \cite[1.A.1]{shaked}, we can show that $P_{out}$ given by (\ref{approx2}), evaluated using the parameter $a+\delta$ is greater than $P_{out}$ evaluated using the parameter $a$, for a constant  $\frac{b}{q}\gamma_0$.
Similarly, the opposite is true for parameters $p+\delta$ and $p$, for a constant $\frac{b}{q}\gamma_0$.
\item [I3)] With a decrease in $\mu_{int}$, outage probability decreases (coverage increases) only when there is a proportionate increase in the target SIR $\gamma_0$. This is so because $\mu_{int}$ leads to an increase in $\mathbb E[(E_i |\vec c^H \vec c_i|^2)^2]$ given by (\ref{term2moment}). This in turn decreases  $a$ given by (\ref{a})  and increases $b$ given by (\ref{b}). This implies that apart from a decrease in $a$, the CDF evaluation point $\frac{b}{q}\gamma_0$ also increases. To really guarantee a decrease in the outage probability according to I2, the target SIR $\gamma_0$ has to be increased to maintain a constant $\frac{b}{q}\gamma_0$. 
\item [I4)] A similar inference can be obtained for an increase in $\mathbb E[(E_i |\vec c^H \vec c_i|^2)^2]$, due to an increase in $\eta_{int}$. A change in $\eta_{user}$ and $\mu_{user}$ leads to a change in all the parameters, namely $a$, $p$, $b$ and $q$ of the outage expression. Hence, it is intractable to form similar inferences using I2 for a change in $\eta_{user}$ and $\mu_{user}$. These inferences have been numerically verified, but not shown here due to space constraints. 
\end{itemize}

\section{Numerical Results}
\begin{figure}[!h]
\centering
\includegraphics[scale=0.37]{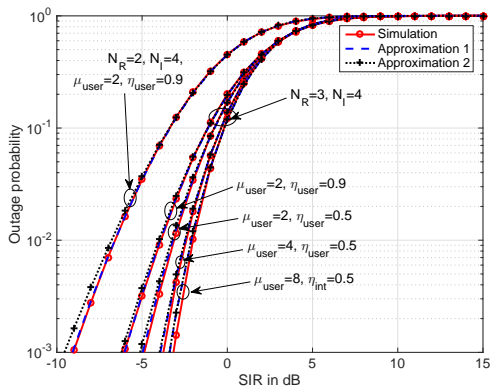}
\caption{Outage probability vs Target SIR for $ \eta_{int}=0.3, \mu_{int}=2, E_I=-1,-3,-5,-7 dB$}
\label{fig1}
\end{figure}

The derived outage probability approximate expression (\ref{Pout}) is verified using Monte-Carlo simulations. For Monte-Carlo simulation, we generate the $\eta-\mu$ random variables $\vec c$ and $\vec c_i$ from the complex pdf given in \cite{eta_mu_phase}. We then use these variables in (\ref{mrcsir1}) to determine the SIR $\gamma$ and plot the CDF of $\gamma$ to obtain the outage probability. From Fig. \ref{fig1}, we can observe that both our approximations are tight for all values of $N_R$ and $N_I$. An increase in $N_R$ results in an increase in the number of independent multi-path and hence leads to increase in SIR. Therefore outage probability decreases as $N_R$ increases, as corroborated by Fig. \ref{fig1}. Increase in $\mu_{user}$ also decreases the outage probability as observed in Fig. \ref{fig1}. This can again be explained by an increase in independent multi-path due to an increase in $\mu$. Similarly, increase in the magnitude of $\eta_{user}$ contributes to increase in the correlation between the real and imaginary components in each cluster of $\eta-\mu$ fading. Hence, an increase in the outage probability can be observed as in Fig. \ref{fig1}. To determine the tightness of the approximations, KL-divergence is calculated empirically between the actual SIR and the approximate SIRs and the same is given in Table \ref{Table1}. We can observe that the approximations become tighter with the interferers becoming close to Rayleigh, i.e.,  $\mu_{int}$ approaching unity and $\eta_{int}$ approaching $0$. This is along expected lines because the approximations are exact for Rayleigh faded interferers. Also, the approximations are tighter for larger $\eta_{user}$ and smaller $\mu_{user}$. Overall, the first approximation is tighter than the second approximation.
\begin{table}
\begin{center}
 \begin{tabular}{||c c c  c c c||}
 \hline
$\eta_{int}$ & $\mu_{int}$ & $\eta_{user}$ & $\mu_{user}$ & Approx 1 & Approx 2\\ [0.5ex] 
 \hline\hline
 0.1 & 2 & 0.1 & 2 & 0.0053 & 0.0057 \\ 
 \hline
 0.1 & 2 & 0.1 & 4 & 0.0074 & 0.0076 \\ 
 \hline
 0.1 & 2 & 0.9 & 4 & 0.0055 & 0.0057 \\ 
 \hline
 0.1 & 4 & 0.9 & 4 & 0.0106 & 0.0108 \\ 
 \hline
 0.9 & 4 & 0.9 & 4 & 0.0185 & 0.0186 \\  [1ex] 
 \hline
\end{tabular}
\caption{KL divergence for the approximations for $N_R=3$, $N_I=2$ and $E_I=-1,-1$ dB}  \label{Table1}
\end{center}
\end{table}
\section{Conclusion}
In this paper, we considered receiver diversity system in the presence of multiple unequal power uncorrelated $\eta-\mu$ interferers. Closed-form approximate outage probability expression for the MRC system was derived for the case when the desired user and the interfering signals are subject to $\eta-\mu$ fading. Extensive Monte-Carlo simulations were performed and the approximation matches the simulation results for all $\eta-\mu$ parameters. The effect of the variation of $\eta$ and $\mu$ on the outage probability was also analyzed using tools from stochastic ordering.
\bibliographystyle{IEEEtran}
\bibliography{macros_abbrev,bibfile}
\end{document}